\documentclass[acus]{pac99}
\usepackage{epsfig}

\begin{document}
\title{NONLINEAR ACCELERATOR PROBLEMS VIA WAVELETS:\\
5. MAPS AND DISCRETIZATION VIA WAVELETS}
\author{A.~Fedorova, M.~Zeitlin, IPME, RAS, St.~Petersburg, Russia
  \thanks{ e-mail:
zeitlin@math.ipme.ru}
\thanks{http://www.ipme.ru/zeitlin.html;
        http://www.ipme.nw.ru/zeitlin.html}}
\maketitle

\begin{abstract}
In this series of eight papers  we
present the applications of methods from
wavelet analysis to polynomial approximations for
a number of accelerator physics problems.
In this part we consider the applications of discrete wavelet 
analysis technique to
maps which come from discretization of continuous nonlinear polynomial
problems in accelerator physics. Our main point is generalization of
wavelet analysis which can be applied for both discrete and continuous
cases. We give explicit multiresolution representation for solutions of
discrete problems, which is correct discretization of our representation
of solutions of the corresponding continuous cases.

\end{abstract}
\section{INTRODUCTION}
This is the fifth part of our eight presentations in which we consider
applications of methods from wavelet analysis to nonlinear accelerator
physics problems.
 This is a continuation of our results from [1]-[8],
in which we considered the applications of a number of analytical methods from
nonlinear (local) Fourier analysis, or wavelet analysis, to nonlinear
accelerator physics problems
 both general and with additional structures (Hamiltonian, symplectic
or quasicomplex), chaotic, quasiclassical, quantum. Wavelet analysis is
a relatively novel set of mathematical methods, which gives us a possibility
to work with well-localized bases in functional spaces and with the
general type of operators (differential, integral, pseudodifferential) in
such bases.
In contrast  with parts 1--4 in parts 5--8 we try to take into account
before using power analytical approaches underlying algebraical, geometrical,
topological structures related to kinematical, dynamical and hidden
symmetry of physical problems.

In this paper we consider the applications of discrete wavelet
analysis technique to
maps which come from discretization of continuous nonlinear polynomial
problems in accelerator physics. Our main point is generalization of
wavelet analysis which can be applied for both discrete and continuous
cases. We give explicit multiresolution representation for solutions of
discrete problems, which is correct discretization of our representation
of solutions of the corresponding continuous cases.

 In part 2 we consider symplectic and Lagrangian structures for the case of
discretization of flows by corresponding maps and  in part 3 construction of
corresponding solutions by applications of
generalized wavelet approach which is based on generalization of
  multiresolution analysis
for the case of maps.

\section{Veselov-Marsden Discretization}
Discrete variational principles lead to evolution dynamics analogous to the
Euler-Lagrange equations [9]. Let $Q$ be a configuration space, then a discrete
Lagrangian is a map $L: Q\times Q\to{\bf R}$. usually $L$ is obtained by
approximating the  given Lagrangian. For $N\in N_+$ the action sum is the map
$S: Q^{N+1}\to{\bf R}$ defined by
\begin{equation}
S=\sum_{k=0}^{N-1}L(q_{k+1}, q_k),
\end{equation}
 where
$q_k\in Q$, $k\ge 0$. The action sum is the discrete analog of the action
integral in continuous case. Extremizing $S$ over $q_1,...,q_{N-1}$ with fixing
$q_0, q_N$ we have the discrete Euler-Lagrange equations (DEL):
\begin{equation}
D_2L(q_{k+1}, q_k)+D_1(q_k, q_{q-1})=0,
\end{equation}
for $k=1,...,N-1$.

Let
\begin{equation}
\Phi: Q\times Q\to Q\times Q
\end{equation}
 and
\begin{equation}
\Phi(q_k, q_{k-1})=(q_{k+1}, q_k)
\end{equation}
 is a discrete function (map), then we have for DEL:
\begin{equation}
D_2L\circ\Phi+D_1L=0
\end{equation}
or in coordinates $q^i$ on $Q$ we have DEL
\begin{equation}
\frac{\partial L}{\partial q^i_k}\circ\Phi(q_{k+1},q_k)+
\frac{\partial L}{\partial q_{k+1}^i}(q_{k+1},q_k)=0.
\end{equation}
It is very important that the map $\Phi$ exactly preserves the symplectic form
$\omega$:
\begin{equation}
\omega=\frac{\partial^2 L}{\partial q_k^i\partial q_{k+1}^j}(q_{k+1},q_k){\rm
d}q_k^i\wedge{\rm d}q^j_{k+1}
\end{equation}

\section{Generalized Wavelet Approach}
Our approach to solutions of equations (6) is based on applications of general and very efficient methods
developed by A.~Harten [10], who produced
a "General Framework" for multiresolution representation of discrete data.

It is
based on consideration of basic operators, decimation and prediction, which
connect adjacent resolution levels. These operators are constructed from two
basic blocks: the discretization and reconstruction operators. The former
obtains discrete information from a given continuous functions (flows), and
the latter produces an approximation to those functions, from discrete values,
in the same function space to which the original function belongs.

 A "new
scale" is defined as the information on a given resolution level which cannot
be predicted from discrete information at lower levels. If the discretization
and reconstruction are local operators, the concept of "new scale" is also
local.

The scale coefficients are directly related to the prediction errors,
and thus to the reconstruction procedure. If scale coefficients are small at a
certain location on a given scale, it means that the reconstruction procedure on
that scale gives a proper approximation of the original function at that
particular location.

This approach may be considered as some generalization of standard wavelet
analysis approach. It allows to consider multiresolution decomposition when
usual approach is impossible ($\delta$-functions case). We
demonstrated the discretization of Dirac function by wavelet packets
on Fig.~1 and Fig.~2.

Let $F$ be a linear space of mappings
\begin{equation}\label{eq:Fin}
F\subset \{f|f: X\to Y\},
\end{equation}
where $X,Y$
are linear spaces. Let also $D_k$ be a linear operator
\begin{eqnarray}
&&D_k: f\to\{v^k\},\quad
 v^k=D_kf,\nonumber \\
&&v^k=\{v_i^k\}, \quad v_i^k\in Y.
\end{eqnarray}
This sequence corresponds to $k$ level discretization of $X$.
Let
\begin{equation}\label{eq:Dk}
D_k(F)=V^k={\rm span}\{\eta^k_i\}
\end{equation}
and the coordinates of $v^k\in V^k$ in this basis are
$\hat{v}^k=\{\hat{v}^k_i\}$, $\hat{v}^k\in S^k$:
\begin{equation}
v^k=\sum_i\hat{v}^k_i\eta^k_i,
\end{equation}\label{eq:vk}
$D_k$ is a discretization operator.
Main goal is to design a multiresolution scheme (MR) [10] that applies to all
sequences $s\in S^L$, but corresponds for those sequences $\hat{v}^L\in S^L$,
which are obtained by the discretization (\ref{eq:Fin}).

Since $D_k$ maps $F$ onto $V^k$ then for any $v^k\subset V^k$ there is  at least
one $f$ in $F$ such that $D_kf=v^k$. Such correspondence from $f\in F$ to
$v^k\in V^k$ is reconstruction and the corresponding operator is the
reconstruction operator $R_k$:
\begin{equation}
R_k: V_k\to F, \qquad D_kR_k=I_k,
\end{equation}
where $I_k$ is the identity operator in $V^k$ ($R^k$ is right inverse of $D^k$
in $V^k$).

Given a sequence of discretization $\{D_k\}$ and sequence of the corresponding
reconstruction operators $\{R_k\}$, we define the operators $D_k^{k-1}$ and
$P^k_{k-1}$
\begin{eqnarray}
D_k^{k-1}&=&D_{k-1}R_k: V_k\to V_{k-1}\\
P^k_{k-1}&=&D_kR_{k-1}: V_{k-1}\to V_k\nonumber
\end{eqnarray}
If the set ${D_k}$ in nested [10], then
\begin{equation}
D_k^{k-1}P^k_{k-1}=I_{k-1}
\end{equation}
and we have for any $f\in F$ and any $p\in F$ for which the reconstruction
$R_{k-1}$ is exact:
\begin{eqnarray}\label{eq:DP}
D_k^{k-1}(D_kf)&=&D_{k-1}f\\
P^k_{k-1}(D_{k-1}p)&=&D_kp\nonumber
\end{eqnarray}
Let us consider any $v^L\in V^L$, Then there is $f\in F$ such that
\begin{equation}
v^L=D_Lf,
\end{equation}
and it follows from (\ref{eq:DP}) that the process of successive decimation [10]
\begin{equation}
v^{k-1}=D_k^{k-1}v^k, \qquad k=L,...,1
\end{equation}
yields for all $k$
\begin{equation}
v^k=D_kf
\end{equation}
Thus the problem of prediction, which is associated with the corresponding MR
scheme, can be stated  as a problem of approximation: knowing $D_{k-1}f$,
$f\in F$, find a "good approximation" for $D_k f$.
It is very important that each space $V^L$ has a multiresolution basis
\begin{equation}
\bar{B}_M=\{\bar{\phi}_i^{0,L}\}_i, \{\{\bar{\psi}^{k,L}_j\}_j\}^L_{k=1}
\end{equation}
and that any $v^L\in V^L$ can be written as
\begin{equation}\label{eq:vL}
v^L=\sum_i\hat{v}_i^0\bar{\phi}_i^{0,L}+\sum^L_{k=1}\sum_j
d_j^k\bar{\psi}_j^{k,L},
\end{equation}
where $\{d_j^k\}$ are the $k$ scale coefficients of the associated MR,
$\{\hat{v}_i^0\}$ is defined by (11) with $k=0$.
If $\{D_k\}$ is a nested sequence of discretization [10] and $\{R_k\}$ is any
corresponding sequence of linear reconstruction operators, then we have from
(\ref{eq:vL}) for $v^L=D_Lf$ applying $R_L$:
\begin{equation}\label{eq:RlDl}
R_LD_Lf=\sum_i \hat{f}^0_i\phi^{0,L}_i+\sum_{k=1}^L\sum_jd_j^k\psi_j^{k,L},
\end{equation}
where
\begin{eqnarray}
&&\phi_i^{0,L}=R_L\bar{\phi}_i^{0,L}\in F,\quad
\psi_j^{k,L}=R_L\bar{\psi}_j^{k,L}\in F,\nonumber\\
&&D_0 f=\sum\hat{f}^0_i\eta^0_i.
\end{eqnarray}
When $L\to\infty$ we have sufficient conditions which ensure that the limiting
process $L\to\infty$ in (\ref{eq:RlDl}, 22) yields a multiresolution basis for $F$.
Then, according to (19), (20) we have very useful representation for solutions
of equations (6) or for different maps construction in the form which are a
counterparts for discrete (difference) cases of constructions from parts 1-4.

\begin{figure}[ht]
\centering
\epsfig{file=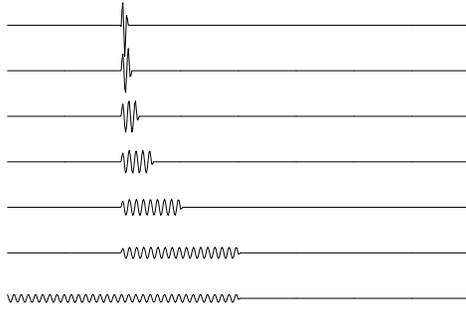, width=82.5mm, bb=0 200 599 590, clip}
\caption{Wavelet packets.}
\end{figure}

\begin{figure}[ht]
\centering
\epsfig{file=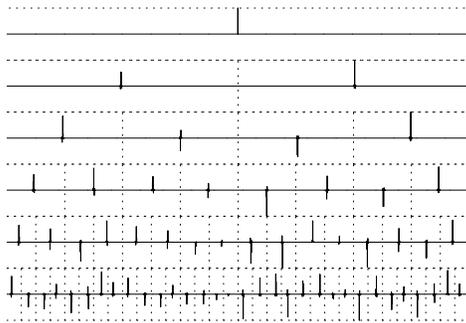, width=82.5mm, bb=0 200 599 590, clip}
\caption{The discretization of Dirac function.}
\end{figure}

We are very grateful to M.~Cornacchia (SLAC),
W.~Her\-r\-man\-nsfeldt (SLAC)
Mrs. J.~Kono (LBL) and
M.~Laraneta (UCLA) for
 their permanent encouragement.

\end{document}